\documentclass[twocolumn,superscriptaddress,preprintnumbers,amsmath,amssymb,pra]{revtex4}
\usepackage{graphicx}
\usepackage{dcolumn,xcolor}
\usepackage{amsmath} 
\usepackage{amssymb}
\usepackage{amsfonts}
\usepackage{bm}
\usepackage[latin1]{inputenc}
\usepackage[dvips]{epsfig}
\usepackage{hyperref}
\usepackage{color}

\usepackage[normalem]{ulem}

\usepackage{ mathrsfs }
\usepackage{wasysym}

\begin{document}

\title{Impact of early-age exposure to power ultrasound\\ on the micromechanical properties of hardened cement paste}

 \author{Martin Chaigne}
 \affiliation{MultiScale Material Science for Energy and Environment, UMI 3466, CNRS-MIT, 77 Massachusetts Avenue, Cambridge, Massachusetts 02139, USA}
 \author{S\'ebastien Manneville}
  \affiliation{MultiScale Material Science for Energy and Environment, UMI 3466, CNRS-MIT, 77 Massachusetts Avenue, Cambridge, Massachusetts 02139, USA}
\affiliation{ENSL, CNRS, Laboratoire de physique, F-69342 Lyon, France}
\affiliation{Institut Universitaire de France (IUF)}
    \author{Michael Haist}
\affiliation{Leibniz University Hannover, Institute for Building Materials, Germany}
\author{Roland J.-M. Pellenq}
  \affiliation{MultiScale Material Science for Energy and Environment, UMI 3466, CNRS-MIT, 77 Massachusetts Avenue, Cambridge, Massachusetts 02139, USA}
  \affiliation{Department of Civil and Environmental Engineering, Massachusetts Institute of Technology, Cambridge, MA 02139}
  \author{Thibaut Divoux}
  \affiliation{MultiScale Material Science for Energy and Environment, UMI 3466, CNRS-MIT, 77 Massachusetts Avenue, Cambridge, Massachusetts 02139, USA}
  \affiliation{ENSL, CNRS, Laboratoire de physique, F-69342 Lyon, France}
  \affiliation{International Research Laboratory, French American Center for Theoretical Science, CNRS, KITP, Santa Barbara, CA 93106, USA}

\date{\today}

\begin{abstract}
Cement paste serves as the universal binder in concrete, formed by mixing water with Ordinary Portland Cement (OPC). In its fresh state, cement paste is a suspension in which the initial particle inventory continuously dissolves, while simultaneously, a growing population of colloidal calcium silica-hydrate (C-S-H) particles forms and progressively builds a percolated structure that densifies and hardens through hydration. Like most colloidal gels, cement paste is sensitive to external stimuli such as mechanical vibrations and temperature variations. Here, we show that power ultrasound (PUS) applied to OPC paste during the very early stages of hydration significantly alters the mechanical properties of the hardened material. Freshly prepared cement pastes are exposed to PUS at varying durations and intensities before curing for 28 days. Micro-indentation testing at the scale of tens of microns reveals that increasing PUS amplitude and prolonged PUS exposure degrade the micro-mechanical properties of hardened cement paste, making it more ductile. This softening behavior evolves continuously with exposure conditions and is consistent with an increase in porosity, likely caused by micro-cracks induced by PUS. This scenario is further supported by micro-scratch testing, which shows higher levels of acoustic emission in PUS-exposed samples. Finally, nano-indentation confirms that the properties of the individual phases composing the hardened paste -- low-density and high-density C-S-H -- remain largely unaffected. These findings offer valuable experimental insights into novel pathways for modifying the mechanical properties of reactive colloidal gels. 
\end{abstract}

\maketitle

\section{Introduction}

Colloidal suspensions are ubiquitous in nature and industry, from blood, milk, and paint to cement paste and cosmetic creams. These systems consist of microscopic particles dispersed in a fluid, where the particle volume fraction and inter-particle forces govern the suspension stability, flow, and overall rheological behavior \cite{Zaccarelli:2009,Joshi:2014,Bonn:2017}. 

When attractive interactions are present, they drive particle aggregation into a volume-spanning network, which imparts solid-like elasto-visco-plastic properties that strongly depend on the mechanical history. External stimuli such as continuous and oscillatory shear \cite{Koumakis:2015,Das:2022,Sudreau:2023,Burger:2024}, electro-magnetic fields \cite{Pauchard:2008,Khatun:2013,Huang:2021}, mechanical vibrations \cite{Nakahara:2005,Kitsunezaki:2016}, and ultrasonic waves \cite{Gibaud:2020,Dages:2021,Ong:2024} can alter particle dispersion, disrupt or enhance aggregation, leading to a variety of microstructures and mechanical properties that may remain ``frozen" when the sample undergoes a rapid liquid-to-solid transition upon cessation of the stimuli.
{This phenomenon, referred to as ``memory effects" \cite{Divoux:2024}, is actively investigated in various systems in order to understand how memory can be stored in and read out from a viscoelastic material \cite{Schwen:2020}.} Beyond fundamental interest, controlling the impact of such stimuli is crucial for many applications that involve, e.g.,  designing soft precursors for hard materials with tailored microstructure \cite{Sudreau:2022,Auxois:2024}, optimizing lithium-ion battery cathode processing \cite{Gupta:2025}, tuning food texture  \cite{Mazzanti:2003,Werner:2023}, or controlling crack patterns in drying colloidal suspensions \cite{Nakahara:2006b,Nakahara:2019}. 

Cement pastes are no exception and show strong sensitivity to external mechanical perturbations. These pastes are obtained by the dissolution of cement grains in water, leading to the precipitation of various hydration products, among which colloidal calcium-silicate-hydrates (C-S-H) form a nanoporous, highly cohesive phase \cite{Masoero:2012,DelGado:2014,Bullard:2011,John:2023}. Their aggregation is driven by Coulombic forces and the interlocking of water and ions \cite{Goyal:2021}, resulting in an irreversible liquid-to-stone transition with critical applications in the construction industry \cite{VanDamme:2018}. The final product -- hardened cement paste -- is a solid matrix with a porous microstructure that influences its strength, durability, and transport properties \cite{Zhou:2019,Ioannidou:2016b}. In practice, perturbations imposed on the cement paste during the very early stage of the hydration process, i.e., when the cement paste behaves as a soft viscoelastic material, impact the long-term structural and mechanical properties of the hardened paste \cite{Ioannidou:2016}. 
In particular, shear-induced particle migration, driven by oscillatory flow conditions, can lead to radial concentration gradients in cement slurries, resulting in central accumulation of particles and potential flow blockages during pumping processes \cite{Kabashi:2023}. Additionally, flow-induced migration in fresh concrete has been shown to cause heterogeneities in particle distribution, affecting the material's uniformity and performance \cite{Spangenberg:2012}. Controlled vibration experiments have further confirmed that mechanical agitation alters the spatial distribution of solids in cement paste, reinforcing the link between flow conditions and microstructural inhomogeneity \cite{Ke:2023}.

Within this framework, power ultrasound (PUS) has been envisioned as a tool to modify the hydration of cement paste, and to disperse various additives aimed at improving and/or tuning the properties of cementitious materials \cite{Martinez:2011,Vaitkevivcius:2018,Ganjian:2018,Xiong:2023b,Lisowski:2024}. In practice, PUS refers to the use of high-intensity ultrasonic waves, with intensities typically above $0.1~\rm W/cm^2$ and frequencies between $20~\rm kHz$ and about $100~\rm kHz$ \cite{Yao:2020}. While PUS does not alter the nature of the final hydration products (especially C-S-H), it changes the pore solution composition by promoting the dissolution of calcium and aluminum ions into the pore solution, enhancing, in particular, the formation of amorphous aluminum hydroxide hydrate, as well as calcium hydroxide (CH) and ettringite \cite{Vaitkevivcius:2018,Ehsani:2022,Xiong:2023c}. 
However, the effect of PUS on early ettringite formation appears to depend on the applied acoustic power: low-intensity ultrasound tends to favor ettringite precipitation, while higher intensities can disturb its early formation -- likely due to localized heating and altered ion availability -- thereby affecting the kinetics of early-age hydration and mechanical performance \cite{Ehsani:2022,Xiong:2023c}. 
In some cases, PUS was shown to accelerate the hydration process, and to reduce the air content in cement paste, eventually leading to a denser hardened paste with improved compressive strength at short times, and up to an 18\% increase for late hydration ages, for instance in pastes of 28 days with a water-to-cement ratio of $0.4$, following a $3~\rm min$ exposure to PUS at a frequency of $28~\rm kHz$ and a power of $912~\rm W$ \cite{Remus:2024,Xiong:2024}. 

In mortars, where cement paste is mixed with sand at a sand-to-cement ratio ranging between $1:1$ and $3:1$, PUS results in even more complex effects. For instance, the compressive strength is generally improved at short times, i.e., within a few days after sample preparation \cite{Peters:2017} -- at least for ultrasonic power lower than $300~\rm W$ \cite{Xu:2023}. However, the compressive strength at 28 days was shown to depend in a non-monotonic way on both the sonication time and the ultrasonic power delivered by the probe, with an optimum at about $240~\rm W$ for a $2~\rm min$ exposure to PUS at a frequency of $20~\rm kHz$ \cite{Xu:2023}. Further studies conducted with an ultrasonic bath aiming at comparing sonicated mortars and mortars obtained from sonicated cement pastes suggest that prolonged exposure to PUS enhances compressive strength, with greater effects observed at lower frequencies, when comparing results obtained at $26~\rm kHz$ and $132~\rm kHz$ for fresh mortars exposed to PUS between $2$ and $10~\rm min$  \cite{Ehsani:2023}.

Despite these recent experimental efforts, the impact of PUS on the mechanical properties of hardened cement paste remains unclear. This is due in part to the vast number of control parameters at stake. Variables such as frequency, power, and exposure time, as well as the sonication method (e.g., bath vs. probe), sample volume, and water-to-cement ratio, can all influence the hydration process. Additionally, PUS introduces thermal energy into the system, leading to sample heating and complicating the distinction between mechanical effects and those arising from thermal activation.
Furthermore, most mechanical testing to date has been conducted at the macroscale, primarily on mortar samples, making it difficult to isolate the effects of PUS on hardened cement paste alone.

Here, we focus on the simplest system, a paste made from Ordinary Portland Cement, and experimentally investigate the impact of PUS applied with a probe sonicator operating at a fixed frequency of $20~\rm kHz$. The duration of the exposure is varied between $0$ and $4~\rm min$, along with the PUS amplitude with a probe delivering up to $750~\rm W$. The effects of PUS, applied to the fresh paste, are quantified through micro- and nano-scale mechanical testing of the hardened paste after 28 days, enabling us to probe smaller length scales than those considered in previous studies. 

We demonstrate that exposing a fresh cement paste to PUS results in a hardened cement paste with degraded micro-mechanical properties, including reduced micro-indentation modulus $M$, hardness $H$, and creep modulus $C$. The degradation is more pronounced with increasing exposure duration and PUS amplitude, and we further show that the relevant control parameter is the acoustic energy injected into the paste. Remarkably, all micro-mechanical measurements collapse onto a master curve when plotting the ductility ratio $M/H$ against the creep modulus $C$, highlighting a continuous evolution of the hardened paste microscale properties as induced by PUS-exposure in the fresh state. The enhanced ductility of the PUS-exposed sample is further supported by micro-scratch tests, which reveal a significant increase in acoustic emission with greater PUS exposure. At the nanoscale, however, the mechanical properties of the two principal phases in the hardened paste, namely low-density and high-density C-S-H, remain unaffected, demonstrating that PUS-induced changes are localized at the microscale, occurring at length scales of tens of microns or more, most likely due to cracking.

These findings suggest that, in our experiments, PUS primarily alters the microstructural organization of the C-S-H network rather than altering its intrinsic composition. We propose a plausible mechanism, consistent with previous literature, in which PUS disrupts early ettringite formation through cavitation and microstreaming effects that generate \textit{localized} heating within the fresh cement paste. 
This disturbance may lead to delayed ettringite formation, introducing microstructural defects and microcracks that ultimately result in a more ductile hardened material. 
Overall, our results highlight the crucial role of early-stage mechanical perturbations in shaping the long-term mechanical performance of cementitious materials.

The manuscript is structured as follows: Section~\ref{sec:exp} details the sample preparation process, including the PUS treatment of fresh cement paste, and introduces the mechanical testing methods. In section~\ref{sec:results}, we illustrate the key role of PUS at the microscale while demonstrating its negligible impact at the nanoscale. Finally, in Section~\ref{sec:disc}, we summarize our findings and discuss them in light of recent literature on PUS-induced effects in cement paste.

\section{Experimental}
\label{sec:exp}

\subsection{Sample preparation}
\label{sec:sampleprep}

Cement samples are prepared by mixing $70~\rm g$ of Ordinary Portland Cement (OPC) clinker powder with $42~\rm g$ of water for $3~\rm min$, yielding a water/cement mass ratio of $0.6$. The liquid cement paste is placed in a metallic $250~\rm mL$ beaker and immediately sonicated with a high-intensity ultrasonic processor (Sonics Vibra-Cell, VCX750) equipped with a $13~\rm mm$ (1/2'') probe producing PUS at a fixed frequency of $20~\rm kHz$. The probe is placed in the center of the beaker, at mid-height in the sample. The control parameters are twofold: the PUS duration $t_s$ varied between $0~\rm min$ and $4~\rm min$, and the PUS amplitude $\mathcal A$ varied between 0 and $115~\rm \mu m$. Moreover, to test the role of the way PUS is applied to the cement paste, samples are prepared in duplicate: PUS is applied at the maximum amplitude either continuously or periodically in bursts of $0.5~\rm s$ duration, separated by a rest period of $0.5~\rm s$. The corresponding data are reported as disks ($\CIRCLE$) and crosses ($\times$), respectively, in the figures discussed thereafter in Section~\ref{sec:results}.  Moreover, for $t_s> 1$~min, we also prepare a series of samples placed in a beaker immersed in an ice bath while exposed to continuous PUS in order to probe the role of the temperature increase of the cement paste induced by PUS. The corresponding data are reported as squares ($\blacksquare$) in the figures discussed thereafter in Section~\ref{sec:results}.  Finally, to test the role of the PUS amplitude, a series of cement paste samples is prepared at fixed PUS duration ($t_s=3$~min) with different PUS amplitudes varying over the entire accessible range.

Following exposure to PUS, the cement paste samples are cast in cylindrical molds (diameter $1.3~\rm cm$, length $7~\rm cm$), sealed with Parafilm, and cured in a calcium hydroxide bath at room temperature ($T\simeq 25^\circ\rm C$) for $9$ days. Cylinders of hardened cement paste are then removed from the mold and stored in sealed plastic bags. After $28$ days, the cylinders are cut into disks of about $1~\rm cm$ thickness using a diamond saw (IsoMet, Buehler). The surface of each disk is polished manually using a series of SiC papers (Buehler) to reach a surface roughness of a few hundred nanometers for mechanical testing \cite{Miller:2008}.

\subsection{Mechanical testing}

\subsubsection{Micro- and nano-indentation tests}
 \label{sec:micronanotest}

Hardened cement pastes exposed to various PUS histories are characterized by statistical indentation tests at both the micro- and nano-scales using a Micro Combi Tester (Anton Paar) and an Ultra NanoIndentation Tester (Anton Paar), respectively, equipped with a Berkovich three-sided pyramid probe. Both experimental setups are equipped with a reference probe, which sits on the sample free surface while running the indentation test. Depth is measured relative to the position of the reference, which prevents artifacts related to electrical and thermal drift, as well as mechanical vibrations. 
In micro-indentation experiments, grids of $15\times 15=225$ indentations are conducted with a grid spacing of $300~\rm \mu m$, hence covering a square area of side $4.2~\rm mm$. In nano-indentation experiments, grids of $21 \times 21=441$ indentations are conducted with a grid spacing of $10~\rm \mu m$, hence covering a square area of side $200~\rm \mu m$. The spacing values are chosen to ensure that two successive indents are independent. 

Each indentation test consists in measuring the indentation depth of the indenter resulting from a symmetric load profile composed of a ramp of increasing load up to a constant value $P_{\rm max}$ reached in {$6~\rm s$}, and maintained for $180~\rm s$ to probe the creep response of the sample, followed by a ramp of decreasing load at a rate identical to the loading part. The maximum load $P_{\rm max}$ is chosen to produce indentation depths of approximately $h_{\rm m} \simeq 17~\rm \mu m$ in micro-indentation and $h_{\rm m} \simeq 290~\rm nm$ in nano-indentation, probing the \textit{local} mechanical properties of the hardened cement paste within a volume of about $50-100~\rm \mu m^3$, and $1~\rm \mu m^3$, respectively. For each indent, the hardness $H$ and the indentation modulus $M$ are determined by the method of Oliver and Pharr \cite{Oliver:1992,Oliver:2004}, while the creep modulus $C$, which corresponds to the inverse of a creep rate, was determined from the contact creep compliance \cite{Vandamme:2012}, following the method previously introduced in Refs.~\cite{Vandamme:2009,Vandamme:2013}. For each sample under study, an indentation grid yields a set of either $225$ or $441$ triplets ($H$, $M$, $C$), which are reported and discussed below in Section~\ref{sec:results}. It is important to emphasize that, while micro-indentation allows us to quantify the average mesoscale properties of the hardened cement paste, nano-indentation allows us to map the mechanical properties of the individual phases composing the cement paste \cite{Constantinides:2006,Randall:2009,Haist:2021} -- see discussion in Section~\ref{sec:nano}. In the latter case, the hardened cement paste is idealized as an assembly of spherical nano-particles \cite{Jennings:2000}, whose properties are driven by cohesion, friction, and packing density \cite{Ulm:2007,Cariou:2008,Haist:2021}.

\subsubsection{Microscratch tests}
\label{sec:microcratchtest}

The fracture toughness $K_c$ of the hardened cement paste is determined at the mesoscale with a Micro Combi Tester (Anton Paar) equipped with a Rockwell C diamond indenter with a spherical tip of radius $R=200~\rm \mu m$ \cite{Gonczy:2005}. In practice, each micro-scratch is performed over $3~\rm mm$ across the surface of a hardened cement disk, which is glued to a square piece of steel and displaced at constant speed $V=100~\mu\rm m.s^{-1}$. A linearly increasing normal load, ramped from $0$ to $30~\rm N$ in $30~\rm s$, is applied during the scratch while the tangential force $F_t$ and the penetration depth $d$ of the probe in the material are recorded, following the same protocol as discussed in ref.~\cite{Hoover:2015}. Prior to each scratch test, the vertical surface profile of the sample is mapped by applying a constant normal load of $0.03~\rm N$  along the scratch length, which was then taken into account to compensate for any tilt in the sample surface. Following the micro-scratch test, the tangential force $F_t(d)$ is normalized according to the following expression:
${F_t}/{\sqrt{2p(d)A(d)}}$, where $p(d)$ and $A(d)$ denote respectively the probe perimeter and its projected contact area orthogonal to the scratch direction. The scratch probe shape function $p(d)A(d)$ is calibrated following the method introduced in ref.~\cite{Akono:2012} by performing scratch tests on a reference material of known fracture toughness, here a carefully cleaned and dried Lexan 9034 (SABIC Innovative Plastic, MA) sample \cite{Akono:2014}. For sufficiently large penetration depth, i.e., $d/R \gtrsim 0.13$, we observe that the normalized tangential force converges to a constant value (see Section~\ref{sec:mesoscale} and Appendix~\ref{sec:appendixC}), which is known to be a reliable estimate of the fracture toughness of the material \cite{Akono:2011,Akono:2012}. For each sample, we conducted about 15 scratch tests. 

Finally, for each scratch test, acoustic emission (AE) is recorded by a piezoelectric sensor (Vallen VS 150-M) operating in resonance mode with a peak at $150~\rm kHz$, which corresponds to the typical frequency resulting from most cracking events in brittle materials \cite{Stebut:1999}. An electronic circuit amplifies and converts the recorded signal from AC to DC. A capacitor then integrates the detected peak into a signal that slowly decays over one second, making it easily readable, which we directly observe in the results provided by the software. 

\subsubsection{Scanning Electron Microscopy}

Following the micro-scratch tests, the surface of the hardened cement paste disks is coated with a carbon layer approximately $20~\rm nm$ thick and dried under ultrahigh vacuum ($8\cdot 10^{-9}~\rm bar$) for at least $24~\rm h$. The sample surface at the locus of the micro-scratch test is imaged with a scanning electron microscope (SEM, JEOL JXA-8200). SEM backscatter images with dimensions $1024\times 768$~pixels are acquired with a resolution of $3.5~\rm \mu m$ per pixel. The beam voltage, current, and working distance are set between $2$ and $5~\rm kV$, $0.5$ and $1~\rm nA$, and $8$ and $11~\rm mm$, respectively.

\section{Results}
\label{sec:results}

\subsection{Micro-mechanical properties}
\label{sec:mesoscale}

\subsubsection{Impact of PUS duration}
\label{sec:PUSduration}

\begin{figure*}[t!]
\centering
\includegraphics[width=0.95\linewidth]{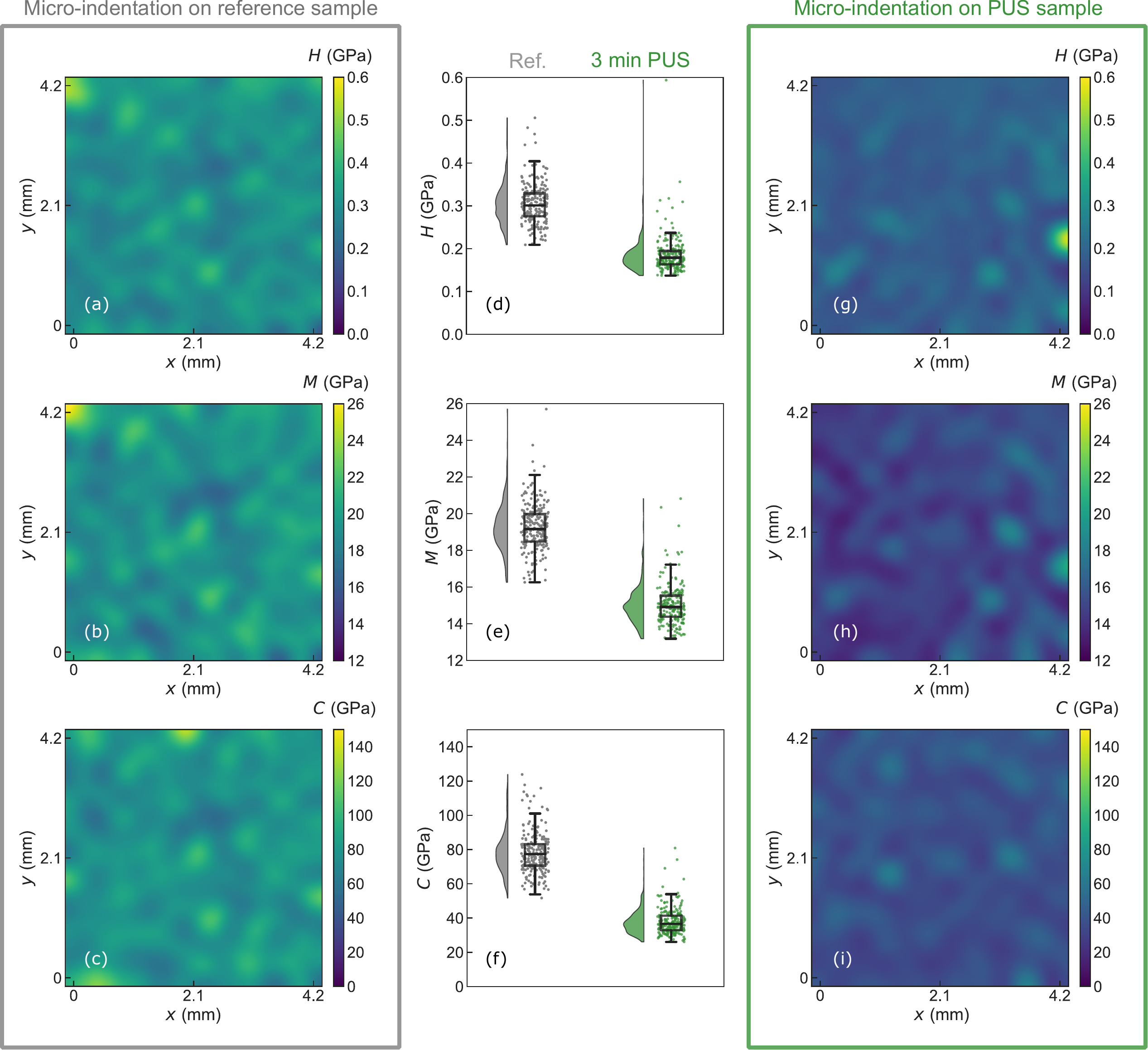}
\caption{Comparison of the micro-mechanical properties of hardened cement paste: hardness ($H$), indentation modulus ($M$), and creep modulus ($C$). Results are shown for a reference sample never exposed to PUS (gray color) and a sample exposed to PUS for $3~\rm min$ in an ice bath before being left to harden (green color). Data were obtained by grid micro-indentation ($15\times 15$ indents with 300~$\mu$m spacing) and are reported in (d)--(f) as ``raincloud'' plots \cite{Allen:2019}. The jittered raw data, represented by the dots, show individual indentations at distinct locations on the sample. The boxplot represents one standard deviation above and below the mean, and the whiskers extend to the limits of the distribution, except for outliers. The side plot shows the probability density function of the data. Corresponding indentation maps are shown respectively in (a)--(c) for the reference sample and (g)--(i) for the PUS-exposed sample. Across all measured properties, the PUS-exposed sample displays degraded mechanical properties compared to the reference sample. 
\label{fig1}}
\end{figure*} 

\begin{figure*}[t!]
\centering
\includegraphics[width=0.95\linewidth]{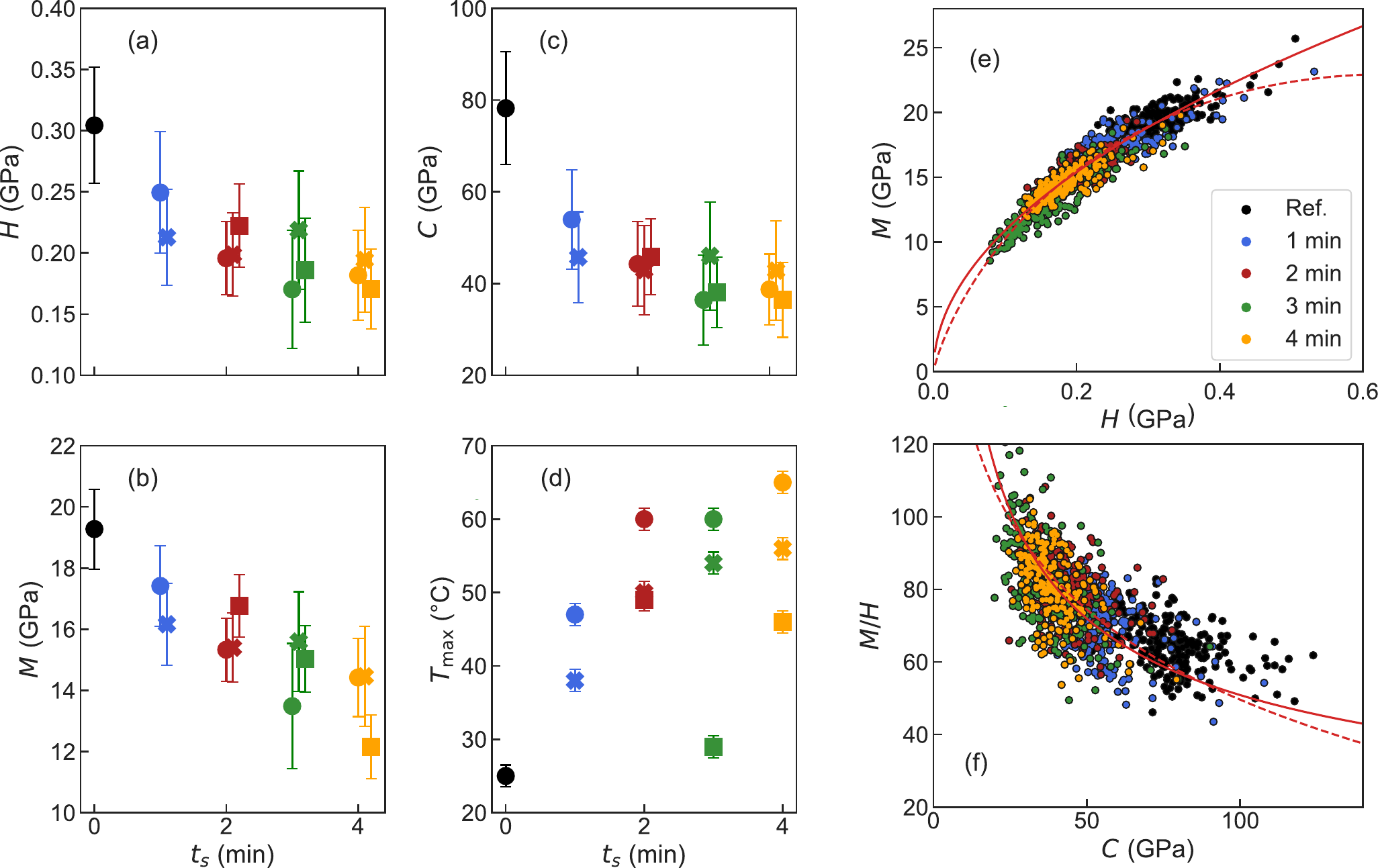}
\caption{Impact of PUS duration $t_s$ on the micro-mechanical properties of hardened cement paste: (a) hardness $H$, (b) indentation modulus $M$, and (c) creep modulus $C$ measured via grid micro-indentation ($15\times 15$ indents with $300~\rm \mu m$ spacing). Experiments were conducted on hardened cement paste prepared following various PUS protocols: continuous PUS at ambient temperature ($\CIRCLE$) or in an ice bath ($\blacksquare$), and burst PUS ($\times$). Error bars represent the standard deviation over about 225 indentations. (d) Maximum temperature reached by the cement paste during PUS exposure.
(e) $M$ vs.~$H$ and (f) ductility ratio $M/H$ vs.~$C$ with data from (a)--(c) shown as scatter plots: a reference sample (\textcolor{black}{$\CIRCLE$}) and four PUS-exposed samples at ambient temperature for $t_s=1$~min (\textcolor{blue}{$\CIRCLE$}), 2~min (\textcolor{red}{$\CIRCLE$}), 3~min (\textcolor{green}{$\CIRCLE$}), and 4~min (\textcolor{yellow}{$\bullet$}), applied in the liquid state.
In (e) and (f), the red curves are the best power-law fit of the data ($M = 34.8 H^{1/2}$ and $M/H=517C^{-1/2}$, respectively). The red dashed curve in (e) is the best fit of the data with the elasto-plastic function from Ref.~\cite{Vandamme:2013}, $M=\alpha H[1-\beta \ln(H/H_0)]$, with parameters {$\alpha=350\pm 4$, $\beta=0.84\pm 0.02$, and $H_0=10^{-4}~\rm GPa$}, the atmospheric pressure. The red dashed curve in (f)  is the same as in (e) under the assumption that $C \simeq H$, as confirmed in Appendix~\ref{sec:appendixA}. Data for the reference and for $t_s=3$~min are the same as in Fig.~\ref{fig1}(d)--(f). 
\label{fig2}}
\end{figure*} 

We first discuss the grid micro-indentation results by comparing the microscale mechanical properties of a reference sample never exposed to PUS, i.e., for $t_s=0$, and a sample exposed to PUS for $t_s=3$~min before being left to harden (see Section~\ref{sec:sampleprep} for detailed sample preparation). Figure~\ref{fig1} illustrates this comparison in the form of raincloud plots for the hardness $H$, the indentation modulus $M$, and the creep modulus $C$. This hybrid representation combines a scatter plot of the raw data, a box plot, and a density plot to provide a comprehensive view of the statistical distribution of each dataset \cite{Allen:2019}. The relevance of such a representation for indentation data was already illustrated on various materials, including nanocomposites \cite{Rao:2022}, sedimentary rocks \cite{Du:2023}, and soft biomaterials \cite{Ciccone:2022}.

First, Fig.~\ref{fig1} shows that the reference hardened cement paste exhibits homogeneous mechanical properties at the mesoscale, well described by Gaussian distributions, with the following average values and standard deviations: hardness $H = 0.30\pm 0.05~\rm  GPa$, indentation modulus $M =19.3\pm 1.3~\rm GPa$, and creep modulus $C=78\pm 12~\rm GPa$. These values are compatible with experimental results from the literature obtained on similar OPC, for the same water-to-cement ratio, as well as predictions for $M$ from homogenization theory \cite{Sanahuja:2007}. 
Comparatively, the sample exposed to $3~\rm min$ of PUS shows significantly degraded mechanical properties with a narrower distribution, i.e., $H =0.19 \pm 0.04~\rm GPa$, $M =15.0\pm 1.1~\rm GPa$, and $C =38 \pm 8~\rm GPa$. Therefore, exposing cement paste to PUS at an early age in the liquid state clearly affects the properties of the hardened paste measured 28 days after PUS exposure.

To illustrate the robustness of our results, we examine the results obtained on samples exposed to PUS for durations varying between 1 and $4~\rm min$. Figure~\ref{fig2}(a)-(c) displays the evolution of the mechanical properties $H$, $M$, and $C$ as a function of the exposure time $t_s$ to PUS. For each exposure time, the experiment is repeated under various conditions, i.e., continuous vs burst PUS (see Section~\ref{sec:sampleprep} for details), yielding consistent and reproducible results, regardless of the PUS application method. All three quantities display a roughly linear decrease for increasing exposure to PUS. Note that a decrease in creep modulus implies a larger creep deformation (or higher creep rate) of the material under the same applied stress. Additionally, Fig.~\ref{fig2}(d) reports the maximum temperature $T_{\rm max}$ reached by the cement paste, while applying PUS. The reference cement paste is prepared at room temperature, i.e., $T \simeq 25^{\circ}$C, while exposure to PUS leads to a temperature increase that varies between $40^{\circ}$C and $60^{\circ}$C. This temperature increase is less pronounced when the PUS treatment is performed on a sample placed in an ice bath [see square symbols ($\blacksquare$) in Fig.~\ref{fig2}(d)]; however, this has no significant impact on the values of $H$, $M$, and $C$, suggesting that the overall temperature increase does not play a key role in the degradation of the micro-mechanical properties -- see discussion in Section~\ref{sec:disc}. 

The same micro-indentation data can also be represented by reporting the scatter plot of the elastic response against the resistance to plastic deformation as $M$ vs. $H$ [Fig.~\ref{fig2}(e)]. In this representation, each PUS exposure time corresponds to a distinct cloud of points. As the duration of PUS exposure increases, the center of mass of the cloud systematically shifts towards lower values of both $H$ and $M$, following a continuous trend. This behavior indicates a progressive alteration of the microstructure in the hardened cement paste induced by early-age exposure to PUS. Notably, the data corresponding to all exposure durations collapse onto a single master curve, which can be interpreted through two complementary frameworks. 
First, the data $M(H)$ is well described by a square-root function shown as a red curve in Fig.~\ref{fig2}(e). This implies that the \textit{recovery resistance} $R_s$, an indicator of energy dissipation during indentation defined as $R_s = 2.263 \cdot M^2/H$ for a Berkovich indenter \cite{Bao:2004}, remains constant across all exposure conditions, with $R_s \simeq 2600~\rm GPa$. In other words, the capacity of the hardened cement paste to dissipate energy through plastic deformation appears insensitive to early-age PUS exposure within the range of conditions investigated.

Alternatively, the $M(H)$ relationship can also be described by the empirical expression $M=\alpha H [1-\beta \ln(H/H_0)]$, which originates from compression experiments in soil mechanics. This expression has been validated for both clays and hardened cement paste -- materials that share a granular microstructure, whose plastic response is governed by changes in packing density \cite{Vandamme:2013}. The logarithmic term arises from the time-dependent creep behavior characteristic of granular materials under compression, which follows a robust logarithmic evolution regardless of the properties of individual grains \cite{Vandamme:2009,Miksic:2013,Srivastava:2017}. This elasto-plastic function fits the experimental data from PUS-exposed samples as well as the square-root law discussed above, at least over the accessible range of $H$ [see the dashed red curve in Fig.~\ref{fig2}(e)]. This observation suggests that PUS exposure primarily affects the packing density of the microstructure, which progressively decreases with increasing exposure duration.

\begin{figure}[ht!]
\centering
\includegraphics[width=0.95\linewidth]{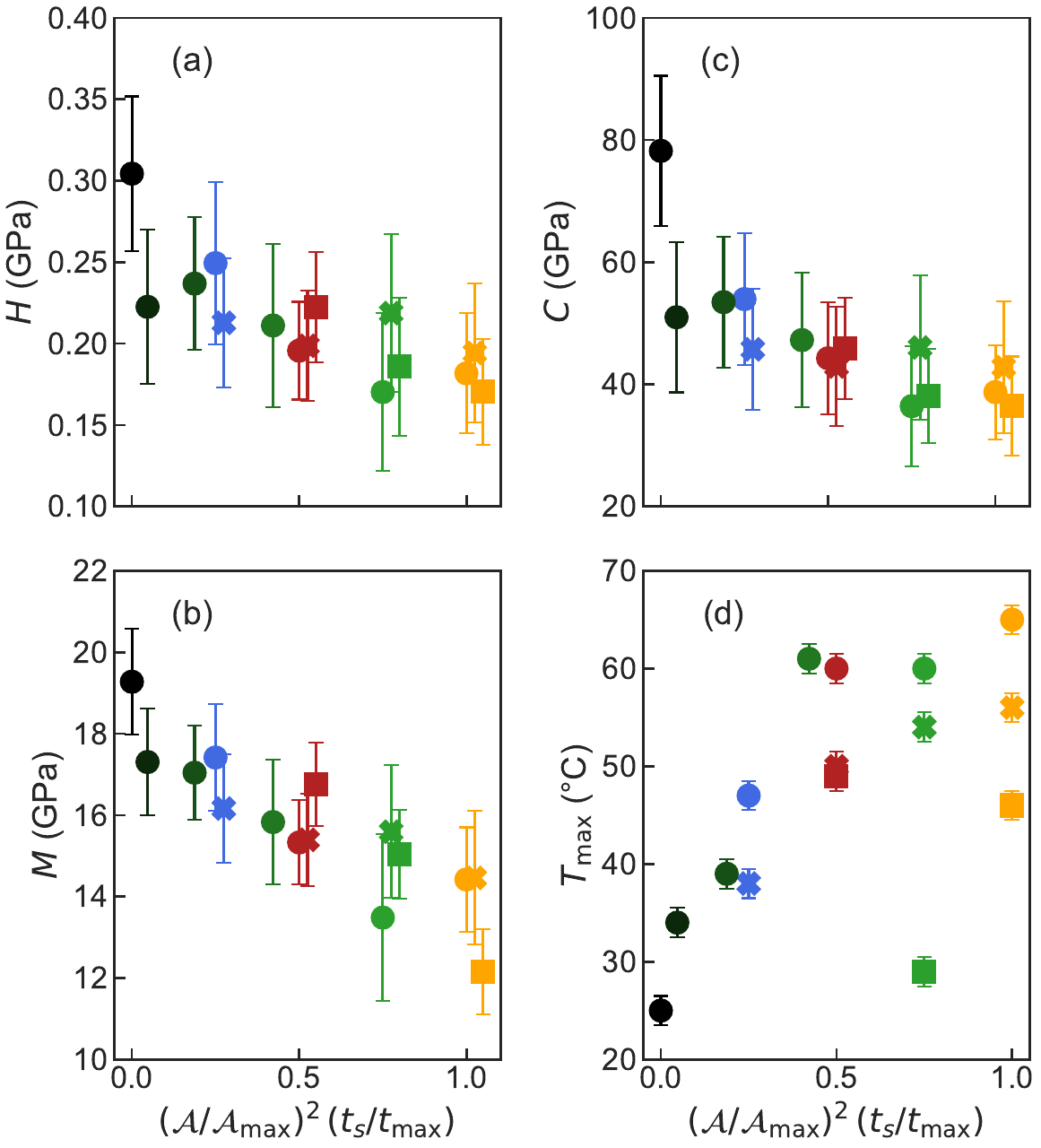}
\caption{Micro-mechanical properties of hardened cement paste, (a) hardness $H$, (b) indentation modulus $M$, and (c) creep modulus $C$, and (d) maximum temperature $T_{\rm max}$ reached by the cement paste during PUS exposure, as a function of a normalized PUS energy $(\mathcal{A}/\mathcal{A}_{\rm max})^2\cdot (t_s/4)$, where $t_s$ is expressed in minutes. Experiments were conducted on hardened cement paste exposed at early age to various PUS protocols: continuous PUS at ambient temperature ($\CIRCLE$) or in an ice bath ($\blacksquare$), and square-wave PUS ($\times$) for various sonication amplitudes $A$ and durations $t_s$. Error bars represent the standard deviation over about 225 indentations. Same data as in Figs.~\ref{fig2}(a)-(d) in the main text, and Figs.\ref{fig4}(a)-(c) in Appendix~\ref{sec:appendixB}.
\label{figA1}}
\end{figure} 

\begin{figure*}[t!]
\centering
\includegraphics[width=0.9\linewidth]{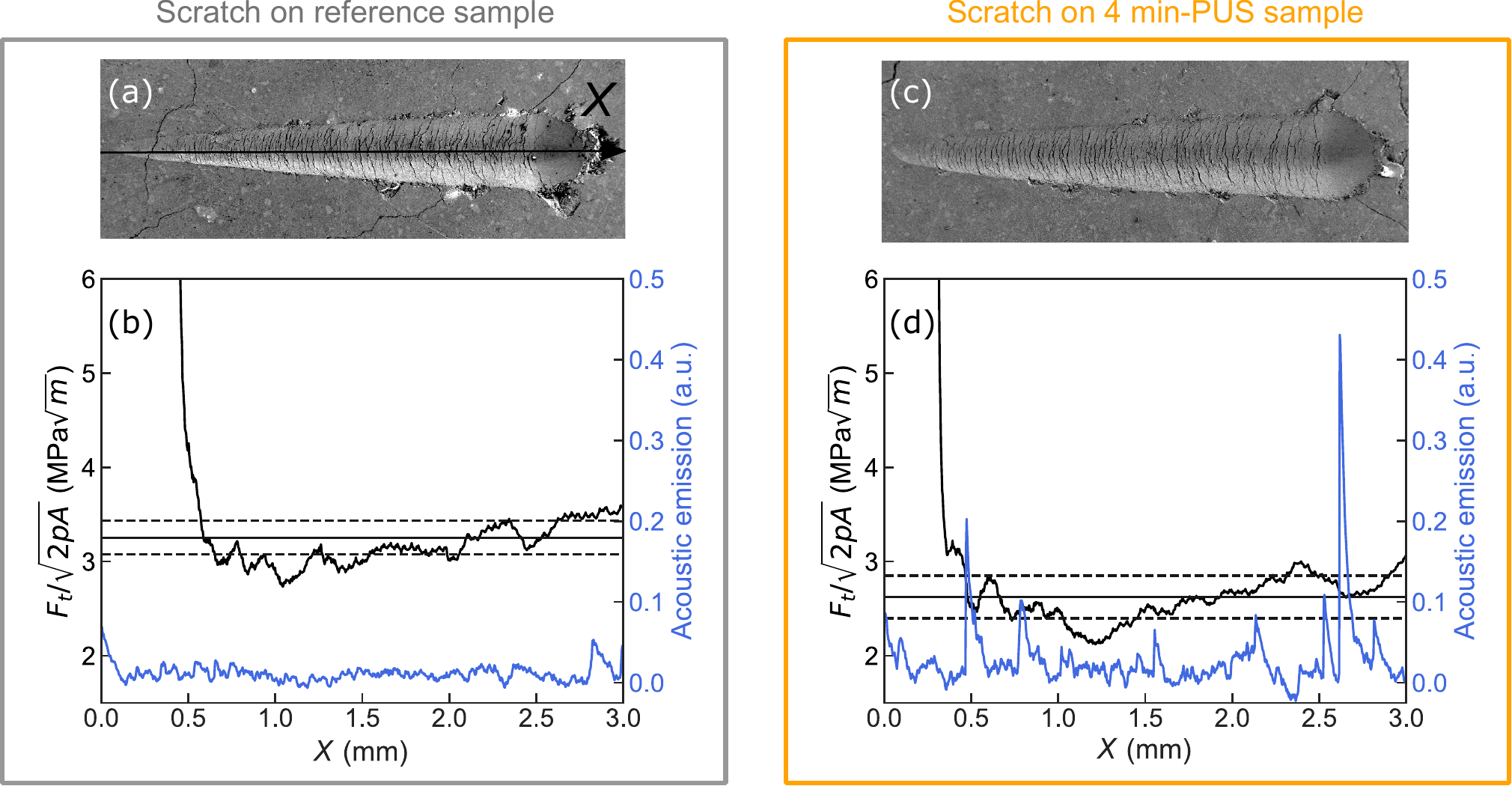}
\caption{Results of scratch tests (a,b) on a reference hardened cement paste, and (c,d) on a hardened paste exposed to continuous PUS for $t_s=4$~min. (a,c) Representative SEM images of a scratched region with dimension $968\times 333$~pixels. The scale is set by the scratch length of 3~mm. (b,d) Normalized tangential force $F_t/\sqrt{2pA}$ (black curve), where $p$ denotes the probe perimeter and $A$ the projected contact area of the probe orthogonal to the scratch direction, and acoustic emission (blue curve) vs.~the position along the scratch length $X$. The horizontal black line shows the average value of $F_t/\sqrt{2pA}$ over the range $0.6<X<3$~mm, which corresponds to the load-independent fracture toughness $K_c$ of the sample. The horizontal dashed lines correspond to the standard deviation. 
\label{fig5}}
\end{figure*} 

Finally, another enlightening representation of the same data set is provided in Fig.~\ref{fig2}(f), where the ductility ratio $M/H$, which corresponds to the inverse of the material yield strain, is reported as a function of the creep modulus $C$. For the reference sample, the ratio $M/H \simeq 60$, which is larger than the value determined by nano-indentation for low-density C-S-H phase  (typically $M/H \simeq 40)$, and much larger than the value of the clinker phase ($M/H \simeq 10$) or that of a purely elastic material, which is approximately $6$ for a Berkovich indenter \cite{Constantinides:2007}. Such a large value indicates the occurrence of plastic deformation during the indentation test. Moreover, the ductility ratio $M/H$, increases from about 60 for the reference sample to 80 for the sample exposed for $4~\rm min$ to PUS, which means that the ductility of the hardened cement paste increases for longer exposure to PUS at early age. In parallel, the creep modulus $C$ decreases from about $80~\rm GPa$ for the reference sample to $40~\rm GPa$ for the sample exposed for $4~\rm min$ to PUS. 
Taken together, these results suggest that cracks form at the mesoscale with increasing PUS duration. As expected, this has a more pronounced effect on hardness than on stiffness, thereby giving the impression of increased ductility. In practice, however, this apparent ductility is unlikely to have any beneficial effect. The assumption of crack formation is further supported by the nanoindentation results presented in Section~\ref{sec:nano}.

\subsubsection{Impact of PUS amplitude and energy}

Here, we examine the impact of varying PUS amplitude while keeping the exposure duration fixed at $t_s=3~\rm min$. The dependence of the key micro-mechanical properties, $H$, $M$, and $C$ are reported in Figure~\ref{fig4} (see Appendix~\ref{sec:appendixB}) as a function of the normalized PUS amplitude, $\mathcal{A}/\mathcal{A}_{\rm max}$, where $\mathcal{A}_{\rm max}=115~\rm \mu m$ is the maximum amplitude achieved by our ultrasonic processor. All three micromechanical properties, both linear and nonlinear, exhibit a continuous decrease with increasing amplitude, consistent with trends observed under conditions of fixed amplitude and increasing exposure duration.

This similarity motivates a unified analysis of both parameters - PUS amplitude and exposure duration. In Figure~\ref{figA1}, we plot the micromechanical properties as a function of $(\mathcal{A}/\mathcal{A}_{\rm max})^2 \cdot (t_s/4)$, a dimensionless quantity which scales with the total acoustic energy delivered to the paste. The resulting data collapse onto a single master curve, indicating that the injected PUS energy is the dominant control parameter governing the mechanical response in our experiments.

\subsubsection{Impact of PUS on fracture toughness}

In addition to hardness and creep moduli,  fracture toughness is another crucial non-linear attribute of the hardened cement paste. Fracture toughness is measured by displacing, at a constant speed, the sample placed in contact with a diamond tip under increasing load (see Section~\ref{sec:microcratchtest} for technical details). The groove left by the scratch test conducted on the reference hardened paste is shown in Fig.~\ref{fig5}(a), while the normalized tangential friction force is reported in Fig.~\ref{fig5}(b) as a function of the coordinate $X$ along the groove. For $X \gtrsim 0.6~\rm mm$, which corresponds to a relative depth for the spherical diamond tip of $d/R \gtrsim 0.13$, the normalized tangential force converges toward a plateau, whose value coincides with the fracture toughness of the sample, here $K_c =3.2 \pm 0.2~ \rm MPa\cdot \sqrt m$. The level of acoustic emission remains negligible across the scratch length. 

The result of the same experiment conducted on a sample after early-age exposure to PUS for $t_s=4~\rm min$ at maximum amplitude is shown in Figs.~\ref{fig5}(c)-(d). The normalized tangential force shows a similar convergence for $X\gtrsim 0.5~\rm mm$, i.e., $d/R \gtrsim 0.13$ towards a value $K_c=2.6\pm0.2~\rm MPa\cdot \sqrt m$ that is slightly lower than that measured on the reference sample. Moreover, the acoustic emission shows strong variations along the scratch with large-amplitude peaks. 
In general, such an acoustic signal is associated with precursors to failure, and the level of acoustic emission depends on the plastic strain. In that context, crack front movements in ductile materials tend to be ``quiet," whereas brittle crack movements tend to be ``noisy" \cite{Pollock:1973}. Our results thus suggest that cement pastes exposed to PUS at an early age display a brittle-like response in the hardened state. However, we have observed from micro-indentation experiments conducted at a similar length scale that the ductility ratio $M/H$ of the hardened cement paste increases with increasing PUS energy, indicating that PUS-exposed samples are more ductile. One way around this paradoxical observation is to assume that defects or pre-existing microcracks are present in the microstructure of PUS-exposed cement paste and are responsible for the higher level of acoustic emission \cite{Scruby:1987}. 

\begin{figure}[t!]
\centering
\includegraphics[width=0.95\linewidth]{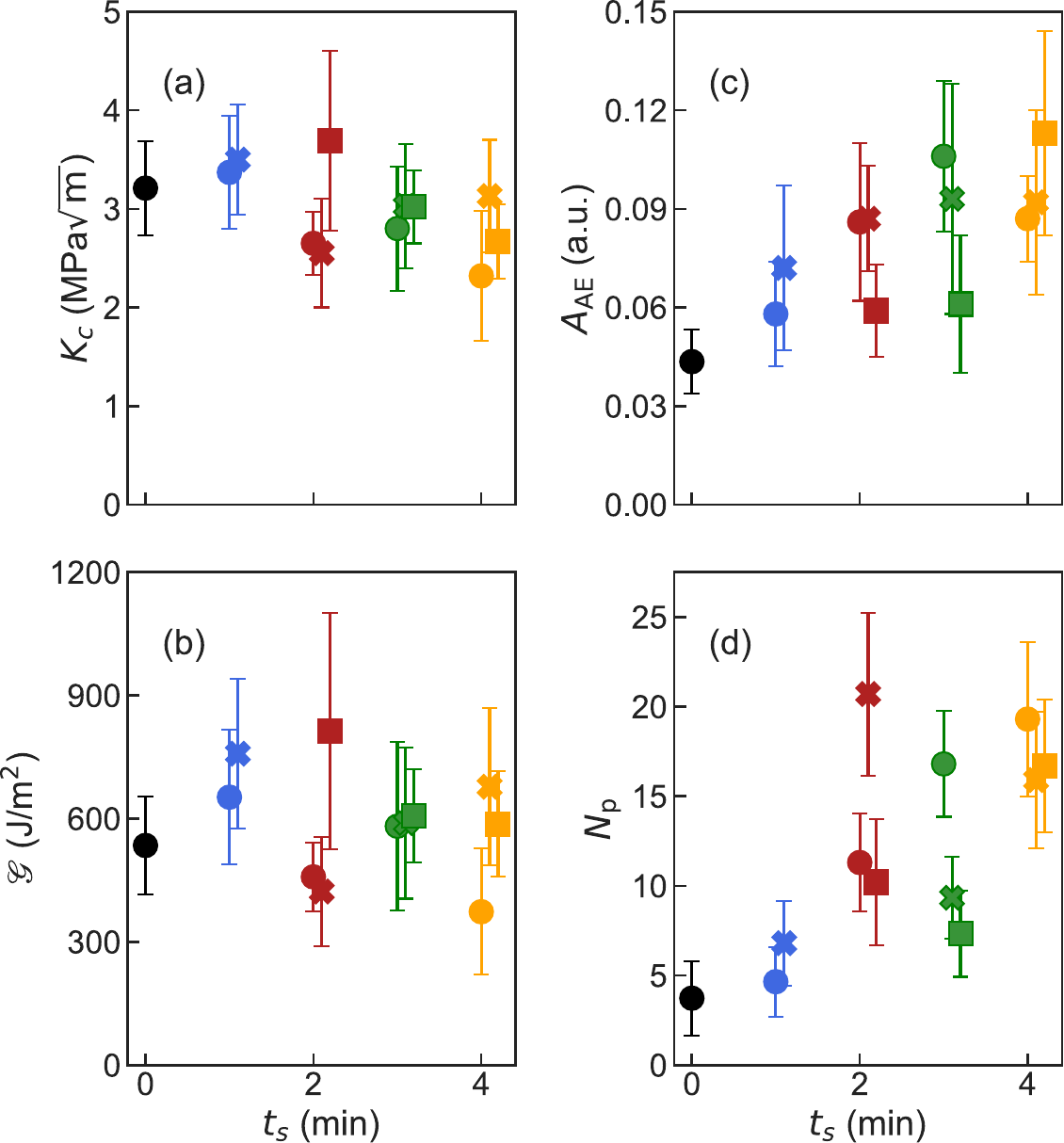}
\caption{Impact of PUS duration $t_s$ on (a) the fracture toughness $K_c$ measured by micro-scratch, (b) the fracture energy $\mathcal{G}=K_c^2/M$, (c) the area computed under the acoustic emission curve, and (d) the number of peaks in the acoustic emission curve observed above a threshold arbitrarily set to 0.04. Experiments were conducted on hardened cement paste exposed at early age to various PUS protocols: continuous PUS at ambient temperature ($\CIRCLE$) or in an ice bath ($\blacksquare$), and burst PUS ($\times$). Error bars correspond to the standard deviation over about 15 micro-scratches. Same color code as in Fig.~\ref{fig2}.
\label{fig6}}
\end{figure} 

\begin{figure}[t!]
\centering
\includegraphics[width=0.95\linewidth]{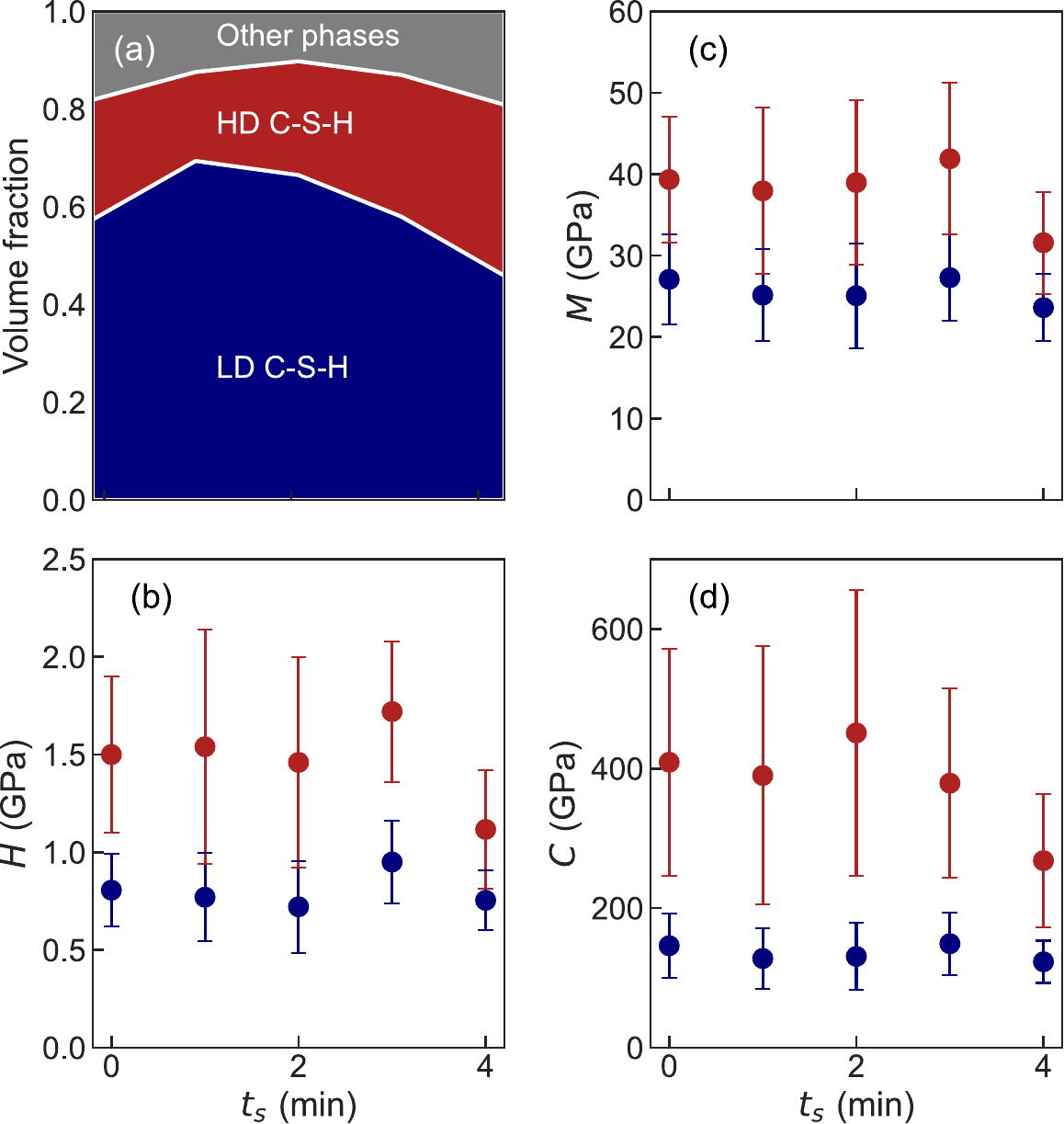}
\caption{Impact of PUS duration $t_s$ on the nanoscale mechanical properties of hardened cement paste: (a) volume fraction of the two main phases, low- and high-density C-S-H, (b) hardness $H$, (c) indentation modulus $M$, and (d) creep modulus $C$. In (b)-(d), the blue (resp. red) color stands for LD- and HD-C-S-H, respectively. Data obtained by grid nano-indentation of $21\times 21$ indents with 10~$\mu$m spacing. Experiments were conducted on hardened cement paste exposed at early age to continuous PUS at ambient temperature. Error bars correspond to the standard deviation of the results. 
\label{fig7}}
\end{figure} 

To further explore the role of PUS exposure on the fracture toughness and acoustic emissions, we conduct micro-scratch tests on hardened cement pastes exposed to various PUS durations. The results are reported in Fig.~\ref{fig6}. We observe that the fracture toughness $K_c$ is only weakly affected by the exposure to ultrasound with $K_c \simeq 3~\rm MPa\cdot \sqrt m$ [Fig.~\ref{fig6}(a)]. The same conclusion holds for the fracture energy, defined as $\mathscr{G}_c=K_c^2/M$ \cite{Irwin:1971}, whose value is roughly constant $\mathscr{G}_c \simeq 59  0~\rm J\cdot m^{-2}$ across all sample tested, irrespective of their exposure time $t_s$ [Fig.~\ref{fig6}(b)]. Note that $\mathscr{G}_c$ can be computed using the modulus $M$ measured in the micro-indentation experiments reported in Section~\ref{sec:PUSduration}, since the typical size probed at each indent, which is about $4$ times the average indentation depth $h_{\rm m} = 17 \pm 1.3~\rm \mu m$ \cite{Ulm:2010}, is comparable with the depth at which the normalized tangential force converged, $d\simeq 0.3 R \simeq 60~\rm \mu m$  \cite{Hoover:2015}. In contrast to $K_c$ and $\mathscr{G}_c$, the acoustic emission quantified by the integral of the acoustic signal and by the total number of peaks occurring during a scratch, shown in Figs.~\ref{fig6}(c) and \ref{fig6}(d), respectively, display a clear increasing trend for increasing exposure duration to PUS. These results confirm that hardened cement pastes, which were exposed to PUS in their fresh state, display an increasing amount of defects or microcracks due to PUS exposure, hence leading to stronger acoustic emission when scratched in their hardened state.

Finally, we consider the plasticity length defined as $\ell=(K_c/H)^2$, which provides an estimate of the characteristic size of plastic deformation \cite{Hoover:2015}. Since $K_c$ is independent of $t_s$ [Fig.~\ref{fig6}(a)], while $H$ decreases for increasing exposure duration to PUS [Fig.~\ref{fig2}(a)], $\ell$ increases for increasing PUS exposure, again mimicking a more ``ductile" behavior, as otherwise shown by the increase in the ductility ratio $M/H$ for increasing exposure to PUS. More specifically, we observe an increase from $\ell \simeq 0.1~\rm mm$ for the reference sample to  $\ell \simeq 0.23~\rm mm$ for $t_s \geq 1$, which then remains constant, irrespective of the duration of ultrasound exposure. 

\subsection{Nano-mechanical properties}
\label{sec:nano}

To determine the impact of PUS on the nanoscale mechanical properties, we conduct nano-indentation tests. We perform nano-indentation grids composed of $21\times 21=441$ indents separated by a distance $D=10~\rm \mu m$, hence covering a square region of interest with a side of $200~\rm \mu m$ (see Section~\ref{sec:micronanotest} for technical details). The maximum load imposed during the load profile was chosen such that the average indentation depth was $h_{\rm m} = 289\pm 40~\rm nm$, thus probing a volume of typical size $4h_{\rm m} \simeq 1.2~\rm \mu m \ll D$, allowing us to treat the indents as independent measurements \cite{Constantinides:2006}. For each sample, we obtain $441$ triplets (H, M, C) that are analyzed by Gaussian mixture modeling as previously reported in Ref.~\cite{Krakowiak:2015}. A Bayesian information criterion is then used to determine the most likely number of phases. The analysis yields between three and five phases, namely low-density C-S-H (LD-C-S-H), high-density C-S-H (HD-C-S-H), calcium hydroxide (CH), mixed phases, and unhydrated clinker \cite{Haist:2021}. Here, we focus only on the LD- and HD-C-S-H phases; the other phases are grouped together as ``other phases." The results are summarized in Fig.~\ref{fig7}, where we report the volume fractions [Figs.~\ref{fig7}(a)] as well as the individual properties of the LD- and HD-C-S-H phases as a function of the exposure time to PUS [Figs.~\ref{fig7}(b)-(d)]. Interestingly, the mechanical properties $H$, $M$, and $C$ of the hardened cement paste are in excellent agreement with our previous studies \cite{Constantinides:2007,Haist:2021}, and independent of the sample's exposure time to PUS, except maybe for the sample exposed for $4~\rm min$, which shows lower values. This outlier behavior at $t_s=4~\rm min$ warrants additional experiments at longer exposure times to clarify whether this trend continues. These results demonstrate that, overall, PUS does not significantly affect the nano-scale properties of the hardened cement paste, underlining the previous microscale conclusions. 

We can therefore conclude that the change in packing density suggested by the master curves observed in the micro-indentation data, and the existence of micro-cracks and defects induced by PUS in the sample microstructure suggested by the increasing level of acoustic emission recorded during scratch tests for increasing PUS exposure, occur at a length scale much larger than a micron, i.e., typically at $10~\rm \mu m $ or above.

\section{Discussion and Conclusion}
\label{sec:disc}

Let us now summarize the main findings of this study and propose a scenario for the impact of PUS on the micro-mechanical properties of hardened cement paste. 

We have demonstrated that exposing fresh cement paste to PUS with a probe sonicator leads to a hardened state with degraded linear and non-linear micro-mechanical properties at 28 days, as evidenced by the consistent decrease in $H$, $M$, and $C$ over a broad range of PUS amplitudes and PUS exposure durations. The key outcomes of the present study are twofold. First, we have shown that the key control parameter of the PUS-induced degradation is the acoustic energy injected into the paste. Second, we have demonstrated that all the micro-mechanical properties fall onto master curves when plotted as $M$ vs.~$H$, and as the ductility factor $M/H$ vs.~the creep modulus $C$ [see Figs.~\ref{fig4}(d)-(c) in Appendix~\ref{sec:appendixB}]. The latter master curve, when considered within the framework of soil mechanics \cite{Vandamme:2013}, shows that PUS mainly affects the \textit{local} packing density of the cement paste, resulting in regions of lower packing densities for more intense or prolonged exposure to PUS. In other words, our results suggest a microscopic scenario in which applying PUS to the fresh cement paste would increase the porosity in the hardened state. This scenario is also supported by the increased level of acoustic emission measured during scratch tests for samples subjected to increasing PUS exposure, indicating the PUS-induced growth of defects or microcracks.
Finally, we have shown that increasing PUS exposure at an early age produces hardened cement pastes that are more ductile, as demonstrated by the increase in the ductility ratio $M/H$ and that of the plasticity length $(K_c/H)^2$ with increasing PUS exposure. Within this framework, our results demonstrate that PUS can be utilized as a tool to gradually modify the mechanical properties of a cement paste with a fixed composition, albeit at the expense of significantly degrading its mechanical properties in the hardened state.

Our results clearly contrast with some previous studies from the literature, where PUS is often reported as being beneficial to the macroscopic mechanical properties (e.g., compressive strength) of hardened cement paste, although some of these mechanical properties may depend nonlinearly on the ultrasonic power or PUS exposure duration \cite{Xiong:2023b,Xu:2023}. In that context, what could explain the difference between our results and those reported above?  
One explanation could be linked to the overall increase in temperature of the sample under the effect of ultrasound. Indeed, we have measured that the temperature increases for increasing injected acoustic energy [see Fig.~\ref{figA1}(d)]. Although the PUS-induced temperature increase is relatively short-lived, typically of about a few tens of minutes, it may affect paste hydration. This conclusion is supported by seminal hydration studies performed (in the absence of PUS) at various temperatures and more recent thermodynamic modeling of the hydration process \cite{Kjellsen:1991,Lothenbach:2008,Gallucci:2013}, which confirm that increasing the temperature during the hydration process increases the porosity of the hardened paste, which could negatively impact its mechanical properties. 

However, the increase in temperature alone cannot explain the degradation of the micro-mechanical properties that we observed. First, previous studies that report improved compressive strength in cement-based samples exposed to PUS also observed a temperature increase \cite{Remus:2024}. Second, we have performed experiments on fresh cement paste sonicated in an ice bath to limit the temperature increase [see $\blacksquare$ in Figs.~\ref{fig2}(d) and \ref{figA1}(d)]. After 28 days, these samples exhibit the same level of degradation as those sonicated at ambient temperature, which are exposed to a larger temperature increase. This finding allows us to rule out temperature increase as the sole explanation.

Here, we propose an alternative explanation for the PUS-induced degradation of the micromechanical properties, which is linked to the hydration products, C-S-H and ettringite. 
When applied during the dormant period of cement hydration, PUS can significantly alter the early-stage phase development of cement \cite{Ehsani:2022}. By exfoliating the aluminate-rich gel layer on cement grain surfaces, PUS increases the dissolution of reactive phases, enhancing the availability of $\rm Ca^{2+}$ and $\rm Al^{3+}$ ions and thereby promoting early ettringite formation -- an argument previously used to justify an improved in the compression strength of PUS-exposed cement-based materials \cite{Vaitkevivcius:2018,Shi:2021,Xiong:2024}. However, the mechanical effects of cavitation -- such as shock waves and microstreaming -- can simultaneously disrupt the fragile ettringite structures and interfere with the subsequent nucleation of outer product C-S-H. As a result, while overall hydration proceeds, early microstructural development may be spatially altered or delayed. 

Within this framework, PUS could hypothetically promote conditions favorable to delayed ettringite formation (DEF) \cite{Taylor:2001}. When reforming at a later stage of the hydration process, ettringite could generate expansive internal stresses that exceed the tensile strength of the surrounding matrix, causing local micro-cracks. This scenario is all the more likely because we are using a sonicator probe, which delivers highly \textit{localized} ultrasonic energy and promotes heterogeneous cavitation, resulting in \textit{very locally} high temperatures compatible with DEF ($T \gtrsim 65^\circ \rm C$), and higher than that produced with an ultrasonic bath used in previous studies, which offers a more uniform exposure of the fresh paste. Additionally, the mechanical disruption caused by cavitation could fragment the emerging C-S-H network, leading to fine, poorly connected gel domains intertwined with aluminate and calcium aluminate phases. Such a microstructure would lack cohesive load-bearing pathways, increasing the risk of micro-crack formation and ultimately degrading the micro-mechanical properties of the hardened cement paste.

While our study provides significant insights into how PUS applied to fresh cement paste influences their fate in the hardened state, several questions remain to be addressed. First, we qualitatively observed that the viscosity of the fresh cement paste increased while applying the PUS. The paste displays a more heterogeneous texture, suggesting that rheological measurements simultaneously with the sonication step would be relevant to characterize the impact of PUS at an early age, before hardening.
Second, our scenario points to an increase in porosity in the hardened paste, which should be quantified by performing systematic porosimetry measurements on samples exposed to increasing levels of acoustic energy. 
Third, our study was performed on small-sized samples of a few $100~\rm cm^3$. It remains unclear how the relative size of the probe sonicator to the container impacts the results reported here, and whether the latter can be scaled up to larger system sizes, for instance, by using an array of probe sonicators \cite{Xiong:2023}, in order to employ PUS in the context of construction applications.    

\begin{acknowledgments}
The authors acknowledge fruitful discussions with E. del Gado, K.~Ioannidou, N.~Randall, and F.-J.~Ulm. The authors also thank S.~Taha for assistance with the SEM imaging. The visit of M.C. to MIT has received support from the MIT-France program, and that of S.M. was funded by the joint CNRS/MIT/Aix-Marseille University Laboratory. T.D. acknowledges the support from the Tremplin CNRS Physics program. This research was supported in part by grant NSF PHY-2309135 to the Kavli Institute for Theoretical Physics (KITP).
\end{acknowledgments}

\appendix

\section{Complementary relations between micro-mechanical properties measured at various PUS exposures}
\label{sec:appendixA}

\begin{figure}[b!]
\centering
\includegraphics[width=0.95\linewidth]{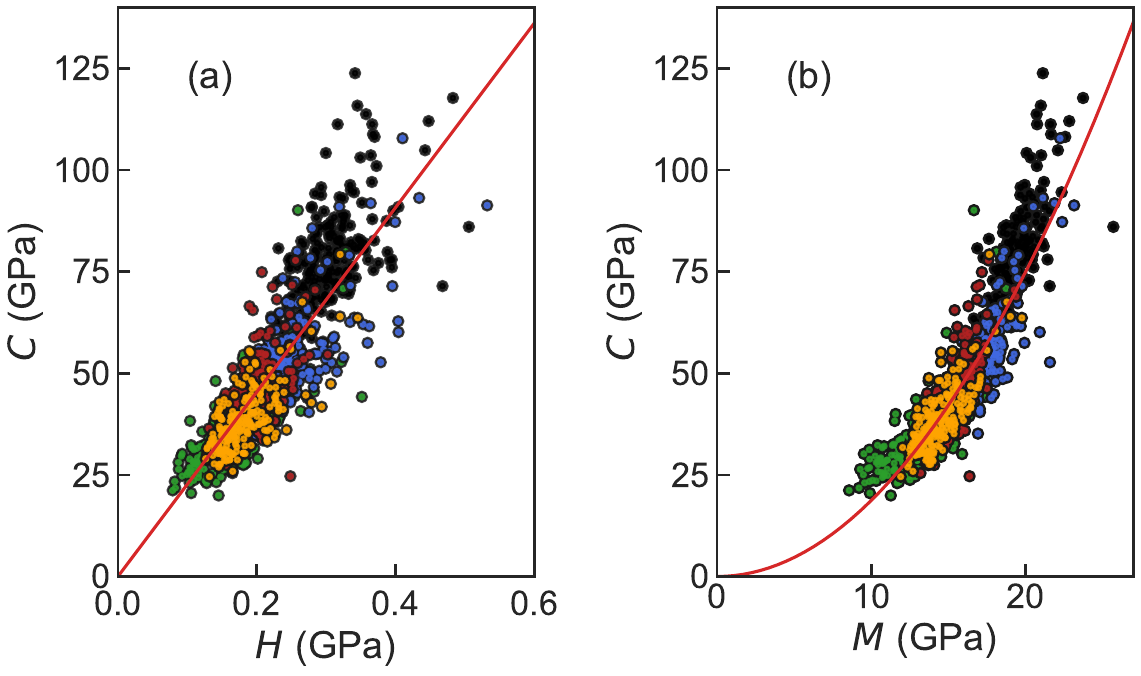}
\caption{Creep modulus $C$ vs (a) Hardness $H$ and (b) indentation modulus $M$. The red curves are the best linear and quadratic fit of the data yielding $C= \lambda H$ and $C=\gamma M^2$, respectively, with $\lambda=227\pm 3$ and $\gamma=0.187\pm0.002\, \rm GPa^{-1}$. Same data as in Figure~\ref{fig2}.  
\label{figA0}}
\end{figure} 

\begin{figure}[bh!]
\centering
\includegraphics[width=\linewidth]{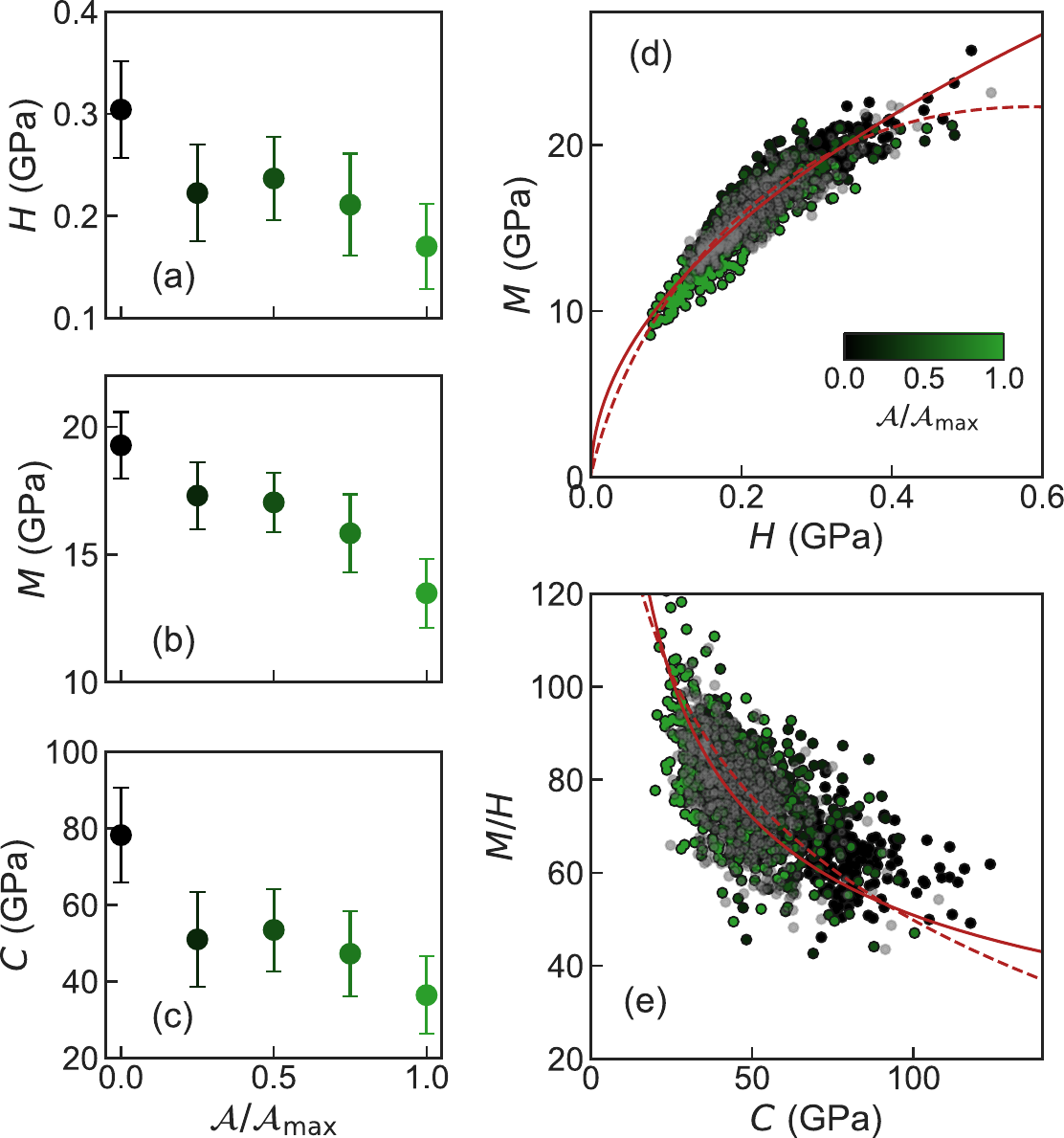}
\caption{Impact of normalized PUS amplitude $\mathcal{A}/\mathcal{A}_{\rm max}$ for a fixed PUS duration ($t_s=3$~min) on the micro-mechanical properties of hardened cement paste: (a)~hardness $H$, (b)~indentation modulus $M$, and (c)~creep modulus $C$. Each data set is obtained by grid micro-indentation ($15\times 15$ indents with $300~\rm \mu m$ spacing). Error bars correspond to the standard deviation of the results over 225 indentations. (d) Indentation modulus $M$ vs.~hardness $H$ and (e) Ductility ratio $M/H$ vs.~creep modulus $C$. Same data as in (a)-(c) in the form of a scattered plot. In addition, the gray points shown in (d) and (e) correspond to the data obtained at various PUS-exposure durations reported in Figs.~\ref{fig2}(e)-(f). The red curves in (d) and (e) are the same as those in Figs.~\ref{fig2}(e) and (f), respectively.
\label{fig4}}
\end{figure} 

Figure~\ref{figA0} shows two additional representations of the data set in Figs.~\ref{fig2}(e) and \ref{fig2}(f), by plotting $C$, respectively, as function of $H$ and as function of~$M$. We find that $C$ is proportional to $H$, while $C$ increases quadratically with $M$. These results are compatible with previous findings from the literature, where $C$ was reported to be proportional to $H$, with a similar prefactor in both nano- and micro-indentation experiments on various types of hardened cement pastes \cite{Vandamme:2009,Vandamme:2013,Stemmermann:2020}, while $C$ was reported to scale as a power-law of $M$ with an exponent of about $1.5$ in nano-indentation tests \cite{Vandamme:2009,Vandamme:2013}. 

\section{Micro-mechanical properties as a function of PUS amplitude}
\label{sec:appendixB}

Figure~\ref{fig4} shows the micro-mechanical properties of hardened cement paste exposed to various PUS amplitudes. Interestingly, the results obtained for various PUS amplitudes collapse onto a master curve when reported as $M$ vs.~$H$ or $M/H$ vs.~$C$. This master curve is identical to the one obtained when considering the data obtained at fixed PUS amplitude for different PUS durations [see gray points in Figs.~\ref{fig4}(c) and ~\ref{fig4}(d)].


\section{Micro-scratch tests: normalized tangential force vs.~relative depth of the scratch probe}
\label{sec:appendixC}

\begin{figure}[ht!]
\centering
\includegraphics[width=0.95\linewidth]{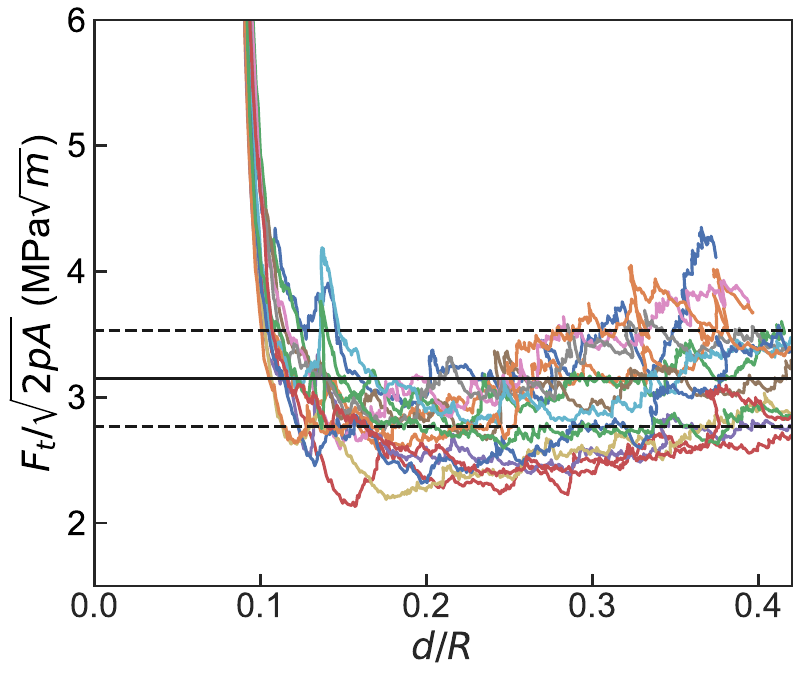}
\caption{Normalized tangential force $F_t/\sqrt{2pA}$ vs.~the relative depth of the scratch probe $d/R$. Each color corresponds to an individual scratch test. The black horizontal line corresponds to the average computed over all measurements, for $d/R \gtrsim 0.13$. The horizontal dashed line corresponds to the standard deviation. Same data as in Fig.~\ref{fig5}(b) in the main text.
\label{figA2}}
\end{figure}

The normalized tangential force, measured for $15$ independent scratch tests performed on the reference hardened cement paste, is shown in Fig.~\ref{figA2} as a function of the normalized scratch depth $d/R$, where $R$ stands for the tip radius of the spherical diamond indenter (see Section~\ref{sec:microcratchtest} for details). In this representation, one can see that for sufficiently large depths, $d/R \gtrsim 0.13$, the normalized tangential force converges to a plateau value, corresponding to the fracture toughness $K_c$. Repeating the test at $15$ different locations allows us to compute a standard deviation for $K_c$, which is shown as horizontal dashed lines in Fig.~\ref{figA2}.

\clearpage


%






\end{document}